\documentclass{nature}
\usepackage{graphicx}
\usepackage{amsmath,color}
\usepackage[final]{pdfpages}

\makeatletter
\let\saved@includegraphics\includegraphics
\AtBeginDocument{\let\includegraphics\saved@includegraphics}
\renewenvironment*{figure}{\@float{figure}}{\end@float}
\makeatother

\title{Quantifying dynamics of failure across science, \\startups, and security}

\author{Yian Yin,$^{1,2}$ Yang Wang,$^{1,3}$ James A. Evans,$^{4,5}$ Dashun Wang$^{1,2,3}$} 

\begin{document}

\maketitle

\begin{affiliations}
 \item Northwestern Institute on Complex Systems, Northwestern University, Evanston, IL 60208, USA
 \item McCormick School of Engineering, Northwestern University, Evanston, IL 60208, USA
 \item Kellogg School of Management, Northwestern University, Evanston, IL 60208, USA
 \item Department of Sociology, University of Chicago, Chicago, IL 60637, USA
 \item Santa Fe Institute, Santa Fe, NM 87501, USA
\end{affiliations}

\begin{abstract}
Human achievements are often preceded by repeated attempts that initially fail, 
yet little is known about the mechanisms governing the dynamics of failure. 
Here, building on the rich literature on innovation\cite{fortunato2018science,azoulay2018toward,harford2011adapt,fleming2001recombinant,wuchty2007increasing,jones2009burden,petersen2012persistence,clauset2015systematic,sinatra2016quantifying,liu2018hot}, human dynamics\cite{jara2018role,hidalgo2015information,barabasi2005origin,gonzalez2008understanding,brockmann2006scaling,castellano2009statistical,malmgren2009universality} and learning\cite{argote1990learning,sitkin1992learning,yelle1979learning,dutton1984treating,shrager1988graph,levitt1988organizational,huber1991organizational,edmondson1999psychological}, we develop a simple one-parameter model that mimics how successful future attempts build on those past. 
Analytically solving this model reveals a phase transition that separates dynamics of failure into regions of stagnation or progression, predicting that near the critical threshold, agents who share similar characteristics and learning strategies may experience fundamentally different outcomes following failures. Below the critical point, we see those who explore disjoint opportunities without a pattern of improvement, and above it, those who exploit incremental refinements to systematically
advance toward success.
The model makes several empirically testable predictions, demonstrating that those who eventually succeed and those who do not may be initially similar, yet are characterized by fundamentally distinct failure dynamics in terms of the efficiency and quality of each subsequent attempt.
We collected large-scale data from three disparate domains, tracing repeated attempts by (i) NIH investigators to fund their research, (ii) innovators to successfully exit their startup ventures, and (iii) terrorist organizations to post casualties in violent attacks, 
finding broadly consistent empirical support across all three domains, which systematically verifies each prediction of our model.
Together, our findings unveil identifiable yet previously unknown early signals that allow us to identify failure dynamics that will lead to ultimate victory or defeat. Given the ubiquitous nature of failures and the paucity of quantitative approaches to understand them, 
these results represent a crucial step toward deeper understanding of the complex dynamics beneath failures, the essential prerequisites for success.
\end{abstract}

Henry Ford went bankrupt twice before founding the Ford Motor Company; 
J.K. Rowling was rejected by twelve publishers before introducing Harry Potter to the world; 
Yet neither came close to
Thomas Edison, who famously failed more than a thousand times before identifying the carbon filament for the light bulb. 
To understand the dynamics of failure, here we collected three large-scale datasets from widely disparate domains (SI S1). 
The first dataset ($D_1$) contains all R01 grant applications ever submitted to the National Institutes of Health (NIH), the world's largest public funder for biomedical research\cite{gross1999relation,ginther2011race,li2015big} (776,721 applications by 139,091 investigators from 1985 to 2015, SI S1.1). 
For each grant application, we obtained ground-truth information on whether or not it was funded, allowing us to reconstruct individual application histories and their repeated attempts to obtain funding. 
Our second dataset ($D_2$) traces start-up investment records from VentureXpert, the official database for National Venture Capital Association\cite{kaplan2016venture} (58,111 startup companies involving 253,579 innovators, SI S1.2). 
Tracing every startup invested by VCs from 1970 to 2016, 
$D_2$ allows us to reconstruct individual career histories counting successive ventures in which they are involved. 
Here we follow prior studies in the entrepreneurship literature\cite{eggers2015dealing,gompers2010performance}, and classify successful ventures as those that achieved initial public offering (IPO) or high value merger and acquisition (M\&A), and correspondingly failed attempts as those that failed to obtain such an exit within five years after their first VC investment.
Going beyond traditional innovation domains, 
we collected our third dataset ($D_3$) 
from the Global Terrorism Database\cite{shapiro2010national}, recording 170,350 terrorist attacks by 3,178 terrorist organizations from 1970 to 2017 (SI S1.3). 
For each organization we trace their attack histories\cite{clauset2012developmental,johnson2011pattern}, and classify success as 
fatal attacks that killed at least one person, and correspondingly failure as those that failed to claim casualties.

Chance\cite{durrett2010probability,bass2011stochastic} and learning\cite{sitkin1992learning,levitt1988organizational} are two primary mechanisms explaining how failures may lead to success. 
If each attempt has a certain likelihood of success, the probability that multiple attempts all lead to failure decreases exponentially with each trial. 
The chance model therefore emphasizes the role of luck, suggesting that success eventually arises from an accumulation of independent trials. 
To test this, we compared the performance of the first and penultimate attempt within failure streaks (SI S4.1), measured by NIH percentile score for a grant application ($D_1$), investment size by VCs to a company ($D_2$), and number of wounded individuals by an attack ($D_3$). 
We find that across all three datasets, the penultimate attempt shows systematically better performance than the initial attempt (Figs.~\ref{hypothesis}c-e, Student's t-test, $p = 1.10\times 10^{-8}~(D_1),~6.01\times 10^{-2}~(D_2),~3.95\times 10^{-5}~(D_3)$). 
Figure 1a rejects that success is simply driven by chance but lends support to the learning mechanism (Fig.~\ref{hypothesis}b), which suggests that failure, and experience more generally, may teach valuable lessons difficult to learn otherwise\cite{sitkin1992learning,argote1990learning,levitt1988organizational,huber1991organizational,argote2012organizational,dahlin2018opportunity}. 
Hence, the more you fail, the more you learn, and the better you perform. 
As such, learning reduces the number of failures required to achieve success, predicting that failure streaks should follow a narrower length distribution (Fig.~\ref{hypothesis}g) than the exponential one predicted by chance (Fig.~\ref{hypothesis}f).
Yet in contrast, across all three domains, failure streak length follows a fat-tailed distribution (Figs.~\ref{hypothesis}h-j, SI S4.2), indicating that despite performance improvement, failures are characterized by longer-than-expected streaks prior to the onset of success. 
Together, these observations demonstrate that neither chance nor learning alone can explain the empirical patterns underlying failures, suggesting that more complex dynamics may be at work.
This raises an intriguing question: What if real settings lie between chance and learning? 

To explore this interplay, 
we develop a simple one-parameter model that in two limiting cases naturally recovers the main predictions of chance and learning  
(Fig.~2, SI S3.1). 
To mimic how future attempts build on previous failures,
we consider that each attempt consists of many distinct components. Take for example the submission of an NIH proposal. Components include constructing a biosketch, assembling a budget, writing a data management plan, adding preliminary data, outlining broad impacts, etc. To simplify our model, here we assume the components are independent and unweighted, with 
each component $i$ being characterized by an evaluation score $x^{(i)}$ (Fig.~\ref{scheme}a).

To formulate a new attempt, one goes through each component, and 
decides to either (1) create a new version (with probability $p$), 
or (2) reuse the best version $x^*$  among the previous $k$ attempts (with probability $1-p$) (Fig.~\ref{scheme}b). 
A new version is assigned a score drawn randomly from a uniform distribution $U[0,1]$, 
approximating the percentile of any score distributions that real systems follow. 
The decision to create a new version is often not random, but driven by the quality of prior versions. 
Indeed, given the best version $x^*$, $1-x^*$ captures the potential to improve it \cite{levitt1988organizational,huber1991organizational}. 
The higher this potential, the more likely one may create a new version, 
prompting us to consider a simple relationship, $p = (1-x^*)^\alpha$, with $\alpha>0$ (SI S3.6). 
Creating a new version takes one unit of time with no certainty that its score will be higher or lower than the previous one. By contrast, reusing the best version from the past saves time, and allows the component to retain its best score $x^*$. 

Here we explore a single parameter $k$ for our model, measuring the number of previous attempts one considers when formulating a new one (Fig.~\ref{scheme}b). Mathematically the dynamical process can be described as
\begin{equation}x_n=
\begin{cases}
U[0,1],~~w.p.~~p\\
x_n^*,~~w.p.~~1-p
\end{cases}
\end{equation}
where $x^*_n= \max\{x_{n-k},\cdots,x_{n-1}\}$.
We quantify the dynamics of the model by calculating (1)
the {\it quality} of the $n$-th attempt, $\left<x_n\right>$, which measures the average score of all components and (2)
the {\it efficiency} after that attempt, $\left<t_n\right>$,  which captures the expected proportion of components updated in new versions.
Let us first consider the two extreme cases:

$k=0$ means each attempt is independent from those past (SI S3.2).  
Here the model recovers the chance model, predicting that as $n$ increases, both $\left<x_n\right>$ and $\left<t_n\right>$ stay constant (Figs.~\ref{scheme}cf). 
That is, without considering past experience, failure does not lead to quality improvement. 
Nor is it more efficient to try again.

The other extreme ($k\rightarrow\infty$) considers all past attempts. 
The model predicts a temporal scaling in failure dynamics (SI S3.3). 
That is, the time it takes to formulate a new attempt decays with $n$, asymptotically following a power law (Fig.~\ref{scheme}h):
\begin{equation}
T_n\equiv\left< t_n\right>/\left< t_1\right>  \sim n^{-\gamma},
\end{equation}
where $\gamma=\gamma_{\infty}=\alpha/(\alpha+1)$ falls between 0 and 1.  
Besides an increased efficiency, new attempts also improve in quality, 
as the average potential for improvement decays following
$\left< 1-x_n\right>  \sim n^{-\eta_{\infty}}$, where $\eta_{\infty}=\min\{\gamma_{\infty},1-\gamma_{\infty}\}$
(Fig.~\ref{scheme}e).
Therefore, in the limit of $k\rightarrow\infty$, our model recovers the canonical result from the learning literature\cite{levy1965adaptation,newell1981mechanisms,anderson1982acquisition,muth1986search,shrager1988graph,mcnerney2011role},
commonly known as Wright's Law\cite{wright1936factors,snoddy1926learning,argote1990learning}. 
This is because, as experience accumulates, high-quality versions are preferentially retained, 
while their lower quality counterparts are more likely to receive updates. 
As fresh attempts improve in quality (Fig.~\ref{scheme}d), they reduce the need to start anew, thus
increasing the efficiency of future attempts (Fig.~\ref{scheme}g).
 
These two limiting cases might lead one to suspect 
a gradual emergence of scaling behavior: 
as we learn from more failures, the scaling exponent $\gamma$ might grow continuously from 0 to $\gamma_{\infty}$. 
On the contrary, as we tune parameter $k$, the scaling exponent follows a discontinuous pattern (Fig.~\ref{solution}a, SI S3.4),
where $\gamma$ only varies within a narrow interval of $\lfloor k^*\rfloor<k<\lceil k^*\rceil+1$ ($k^*\equiv 1/\alpha$).
Indeed, as we increase $k$, agents consider more past experiences.
Yet, when $k$ is small ($k<k^*$), 
the system converges back to the same asymptotic behavior as $k=0$ (Fig.~\ref{solution}abe). 
In this region, although new versions build on past $k$ attempts, 
$k$ is not large enough to retain a good version once it appears. 
As a result, while performance might improve slightly in the first few attempts, it quickly saturates. In this region, agents reject prior attempts and thrash around for new versions, not gaining enough feedback to initiate a pattern of intelligent improvements, prompting us to call it the {\it stagnation} region.
Once $k$ passes the critical threshold $k^*$, however, scaling behavior emerges (Fig.~\ref{solution}acf), 
indicating that the system enters a region of {\it progression}, where failures lead to continuous improvement in both quality and efficiency.
Nevertheless, with a single additional experience considered, 
the system quickly hits the second critical point $k^*+1$, beyond which the scaling exponent $\gamma$ becomes independent of $k$ (Fig.~\ref{solution}adg). 
This means, once $\lceil k^*\rceil+1$ number of prior failures are considered, the system is characterized by the same dynamical behavior as $k\rightarrow\infty$. 
The second critical point indicates that $\lceil k^*\rceil+1$ attempts are sufficient to recover the same rate of improvement as considering \textit{every failure} from the past. 

Most importantly, we show that the two critical points in our model  
can be mapped to phase transitions within a canonical ensemble consisting of three energy levels (Methods, SI S3.5).
The uncovered phase transitions indicate that small variations at the microscopic level may lead to fundamentally different macroscopic behaviors. 
For example, two individuals near the critical point may initially appear identical in their learning strategy or other individual characteristics, yet depending on which region they inhabit, 
their outcomes following failures could differ dramatically (Figs.~\ref{solution}hi). 
In the progression region ($k>k^*$), 
agents exploit rapid refinements to improve through past feedback.
By contrast, those in the stagnation region ($k<k^*$) do not seem to profit from failure, 
as their efforts stall in efficiency and saturate in quality. 
As such, the phase transitions we uncovered in our simple model make four distinct predictions, which we now test directly in the contexts of science, startups, and security.

\textbf{Prediction A}: \emph{Not all failures lead to success}.
While we tend to focus on examples that eventually succeeded following failures, such as Ford, Rowling, and Edison, the stagnation region 
predicts that there exists a non-negligible fraction of cases that do not succeed following failures.
We can test this prediction in our three datasets by measuring the number of failed cases that do not achieve eventual success. 
To eliminate the possibility that such individuals or organizations were simply in the process of formulating their next attempt, 
we focus on cases where it has been at least five years since their last failure. 
We find that, across all three domains, members of the ``non-success'' group not only exist, but their size is of a similar order of magnitude as the success group (Figs.~\ref{empirical}a-c inset).
Interestingly, when we measure the number of consecutive failures for the non-success group, we find that its distribution is statistically indistinguishable from that of the failure streaks (Figs.~\ref{empirical}a-c, Kolmogorov-Smirnov test, $p = 0.286~(D_1),~0.175~(D_2),~0.931~(D_3)$),
indicating that people who ultimately succeeded did not try more or less than their non-successful counterparts.

\textbf{Prediction B}: \emph{Early dynamical signals separate the success group from the non-success group}. 
The model predicts that the success group is characterized by power-law temporal scaling (Eq.~2), which is absent for the non-success group (Fig.~\ref{solution}h). 
Therefore, those who eventually succeed and those who do not may be initially similar but can follow fundamentally different failure dynamics distinguishable at an early stage. 
To test this prediction, we measure the average inter-event time between two failures $T_n$ as a function of the number of failures (SI S4.3). Figures~\ref{empirical}d-f unveil three important observations.\\
(i) For the success group, 
$T_n$ decays with $n$ across all three domains,
approximately following a power law, as captured by (2) (Fig.~\ref{empirical}j, SI S4.3). 
The scaling exponents are within a similar range as those reported in learning curves\cite{dutton1984treating,argote1990learning}, 
further supporting the validity of power law scaling. 
Although the three datasets are among the largest in their respective domains, agents with a large number of failures are exceedingly rare, limiting the range of $n$ that can be measured empirically. 
We therefore test if alternative functions may offer a better fit, finding power law to be the consistently preferred choice (SI S5.2).\\
(ii) The temporal scaling disappears, when we measure the same quantity for the non-success group (Figs.~\ref{empirical}d-f), consistent with predictions of the stagnation region in our model. 
Regression analysis further supports this observation, 
showing that the association between $T_n$ and $n$ is not statistically significant.\\
(iii) The two groups show clearly distinguishable failure dynamics as early as $n=2$ (Student's t-test, $p =4.57\times10^{-3}~(D_1),~7.73\times 10^{-3}~(D_2),~4.99\times 10^{-2}~(D_3)$), demonstrating intriguing early signals that separate those who eventually succeed from those who do not.

The observations uncovered in Figs.~\ref{empirical}d-f are intriguing for two main reasons. 
First, failures captured by the three datasets differ widely in their scope, scale, definition, and temporal resolution,
yet despite these differences, they are characterized by remarkably similar dynamical patterns predicted by our simple model.
Second, membership in the two groups appears to be determined by the last attempt only. 
For example, comparing agents who failed 10 times but succeeded on the 11th with those who gave up after 10 failures,
one might expect that it was the last attempt that separated the two cases.
Yet, as the model predicts, the success and non-success group each follows their respective, highly predictable patterns, 
distinguishable long before the eventual outcome becomes apparent.
Indeed, we use $D_1$ to set up a prediction task (see Methods), to predict the ultimate victory or defeat using only the temporal features, yielding a substantial predictive power. 
Despite the ubiquity of power laws across a wide variety of settings\cite{clauset2009power,barabasi1999emergence,barabasi2005origin,brockmann2006scaling,castellano2009statistical,bettencourt2007growth,gonzalez2008understanding} and the foundational literature on learning curves\cite{wright1936factors,snoddy1926learning,yelle1979learning,newell1981mechanisms,anderson1982acquisition,muth1986search,argote1990learning,shrager1988graph,ritter2001learning,mcnerney2011role} (SI S2), 
none of the existing models, to our knowledge, anticipated the existence of such early signals (Table S1). 
To test if the observed patterns in Figs.~\ref{empirical}d-f may simply reflect preexisting population differences,  
we take agents who experienced a large number of failures (large $n$, hence most different toward the end), and measure their performance during the first attempt. 
We find that for all three domains, the two populations were statistically indistinguishable in their initial performance (Figs.~\ref{empirical}g-i), which leads us to the next prediction: 

\textbf{Prediction C}: \emph{Diverging patterns of performance improvement}. 
Although the two groups may have begun with similar performance, 
the model predicts that they experience different performance gains through failures (Fig.~\ref{solution}i). 
We therefore compared performance at first and second attempts, finding significant performance improvement for the success group (Figs.~\ref{empirical}g-i, $p = 9.28\times10^{-2}~(D_1),~4.18\times 10^{-2}~(D_2),~5.49\times 10^{-3}~(D_3)$), 
which is absent for the non-success group ($p = 0.492~(D_1),~0.219~(D_2),~0.824~(D_3)$).  
We further repeated our measurements by comparing the first and penultimate attempt, and the first and halfway attempt, 
and for both cases, we arrive at the same conclusion (SI S6.3, Fig.~S28). 
This prediction explains the patterns observed in Figs.~\ref{hypothesis}c-e,   
which leads us to the second puzzle raised by Fig.~1: if performance improves, why are failure streaks longer than we expect?

One key difference between the success and non-success groups is their propensity to reuse past components. 
From the perspective of exploration vs.~exploitation\cite{march1991exploration, foster2015tradition}, 
although reuse helps one to retain a good version when it appears, it could also keep one in a suboptimal position for longer.  
Indeed, we analytically calculate the streak length distribution predicted by our model, offering our final prediction:\\~~
\textbf{Prediction D}: \emph{The length of failure streaks follows a Weibull distribution} (Fig.~\ref{empirical}k):  
\begin{equation}
P(N\geq n)\sim e^{-(n/\lambda)^{\beta}}, 
\end{equation}
which explains its fat-tailed nature observed in Figs.~1h-j. Moreover, the shape parameter $\beta$ is connected with the temporal scaling exponent $\gamma$ through a scaling identity (SI S3.8)
\begin{equation}
\beta+\gamma=1.
\end{equation}
This means, if we fit the streak length distribution in Figs.~1h-j to obtain the shape parameter $\beta$ (Fig.~4k), it should relate to the temporal scaling exponent $\gamma$ (Fig.~4j), obtained from Figs.~4d-f. 
Comparing $\beta$ and $\gamma$ measured independently across all three datasets shows consistency between our data and the scaling identity (Eq. 4) (Fig.~4l).

We further test the robustness of our results along several dimensions (SI S6). We vary the definitions of success group (S6.1) by excluding revisions in $D_1$ (Fig.~S20), changing the threshold of high-value M\&As in $D_2$ (Figs.~S24-25), and restricting types of attacks in $D_3$ (Fig.~S26). 
We also vary the definition of non-success groups (S6.2, Figs.~S14-19), and test other measures to approximate performance (S6.3-S6.4, Figs.~S23,S27). 
Across all variations, our conclusions remain the same.

An alternative interpretation for the stalled efficiency of the non-success group is a hedging behavior against failures---their efficiency did not improve because they spent more effort elsewhere. The three professions we studied, ranging from NIH investigators to entrepreneurs to terrorists, involve varied levels of risk, exposure, and commitment, which renders this explanation less likely. Nevertheless, one irony suggested by our model is that agents in the stagnation region did not work less. Rather they made more, albeit unnecessary modifications to what were otherwise advantageous experiences. 

The model also offers relevant insights for the understanding of learning curves. For example, the second critical point of the model suggests the existence of a minimum number of failures one needs to consider ($k^*+1$), indicating that it is not necessary to learn from {\it all} past experiences to achieve a maximal learning rate. This finding poses a potential explanation for the widespread nature of Wright's law across a wide variety of domains, particularly given the fact that in many of those domains not all past experiences can be considered (SI S2).

The one-parameter model developed here represents a minimal model (SI S3.7), which can be further extended into richer frameworks with more flexible assumptions. For example, $\alpha$ captures the propensity to change given feedback, and so can be leveraged to incorporate population heterogeneity into the model, pointing to promising future research that explores the interplay between $\alpha$ and $k$ parameters. Further, the assumed relationship between $p$ and $(1-x^*)$ is not limited to a power law but can be relaxed into its asymptotic form. Indeed, we show that, as long as the function satisfies $\frac{\ln p}{\ln (1-x^*)}\rightarrow \alpha$ as $x^*\rightarrow 1$, the model offers the same predictions\cite{muth1986search} (SI S3.6). Lastly, as a simple model, it does not take into account many of the complexities in real settings that may affect failure dynamics, such as knowledge depreciation\cite{arbesman2013half}, forgetting and transfer\cite{argote2012organizational} or vicarious learning from others\cite{madsen2010failing}. Despite its simplicity, the model accurately predicts several fundamental patterns governing the dynamics of failure. 
As such, it also offers a theoretical basis, where additional factors can be incorporated, including individual and organizational characteristics that may affect eventual success and failure outcomes\cite{argote1990learning,cannon2005failing}.

Together, these results support the hypothesis that if future attempts systematically build on past failures, the dynamics of repeated failures may reveal statistical signatures discernible at an early stage. Traditionally the main distinction between ultimate victory and defeat following failure has been attributed to differences in luck, learning strategies or individual characteristics, but here our model offers an important new explanation with crucial implications: Even in the {\it absence} of distinguishing initial characteristics, agents may experience fundamentally different outcomes.
Indeed, Thomas Edison once said, `\textit{Many of life's failures are people who did not realize how close they were to success when they gave up}.' Our results unveil identifiable early signals that help us predict and anticipate the eventual victory or defeat to which failures lead. Together, they not only deepen our understanding of the complex dynamics beneath failure, they also hold lessons for individuals and organizations that experience failure and the institutions that aim to facilitate or hinder their eventual breakthrough. 
\clearpage

\begin{methods}
\subsection{Phase transitions.}
To understand the nature of two transition points of our model, we consider a canonical ensemble of $N$ particles ($N\rightarrow\infty$) and three energy states $E_a(h) = 1$, $E_b(h) = (2h-1)^2$, and $E_c(h) = 1$, where $h$ denotes the external field. We can write down the partition function of the system $Z = e^{-N E_a(h)}+ e^{- N E_b(h)} + e^{- N E_c(h)}$, and calculate its free energy density $f = \ln Z/N$. In this system, it can be shown that the magnetization density $m=\frac{df}{dh}$ is discontinuous at the boundary of two energy states $E_a(h)=E_b(h)$ and $E_b(h)=E_c(h)$, characterized by two phase transitions at $h=0$ and $h=1$, respectively.

We notice that the canonical ensemble considered above has a one-to-one mapping to our model. Indeed, denoting with $\Gamma\equiv k^*\gamma/(1-\gamma)$ and $K\equiv k-k^*$, we can rescale the system as $\Gamma=\min\{\max\{\Gamma_a(K),\Gamma_b(K)\},\Gamma_c(K)\}$, where $\Gamma_a(K)=0$, $\Gamma_b(K)=K$, and $\Gamma_c(K)=1$, allowing us to map the two systems through $f\rightarrow (2\Gamma-1)^2$, $N\rightarrow\ln n$, $h\rightarrow K$, and $E_i(h)=[2\Gamma_i^2(K)-1]^2$ (Fig.~S8).

To understand the origin of the two transition points, 
we can calculate the expected life span of a high-quality version, obtaining $\langle u(x)\rangle\sim \langle (1-x)^{-\min\{k/k^*,1/k^*+1\}}\rangle$ (SI S3.4). The first critical point $k^*$ occurs when the first moment $\langle u\rangle$ diverges. Indeed, when $k$ is small ($k<k^*$), $\langle u\rangle$ is finite, indicating high-quality versions can only be reused for a limited period. Once $k$ passes the critical point $k^*$, however, $\langle u\rangle$ diverges, offering the possibility for a high-quality version to be retained for an unlimited period of time. The second critical point arises due to the competition between two dynamical forces: (a) whether the current best version gets forgotten after $k$ consecutive attempts in creating new versions (dominated by the $k/k^*$ term); or (b) it is substituted by an even better version (dominated by the $1/k^*+1$ term).

\subsection{Predicting ultimate success.}

We use a simple logistic model to predict whether one may achieve success following $N$ previously failed attempts in $D_1$, 
using only temporal features $t_n$ ($1\leq n\leq N-1$) as predictors.
To evaluate prediction accuracy, we calculate the AUC curve 
over 10-fold cross validation. 
We find that, by observing timing of the first three failures alone, our simple temporal feature yields high accuracy in predicting the eventual outcome with an AUC close to 0.7, significantly higher than random guessing (Mann-Whitney rank test, $p<10^{-180}$, SI S5.1, Fig.~S10a). We repeated the same prediction task on $D_2$ and $D_3$, arriving at similar conclusions (SI S5.1, Fig.~S10). 
The predictive power documented here is somewhat unexpected. 
Indeed, there are a large number of documented factors that affect the outcome of a grant application\cite{ginther2011race,boudreau2016looking,bromham2016interdisciplinary,banal2016key,ma2015anatomy}, ranging from prior success rate to publication and citation records to race and ethnicity of the applicant. 
Yet here we have blatantly ignored these factors, using only features pertaining to temporal scaling as prescribed by our model.
This suggests that this predictive power represents a lower-bound, which could be further improved and leveraged by incorporating additional factors.

\end{methods}

\begin{addendum}
 \item The authors wish to thank Chaoming Song, Brain Uzzi, and Eli Finkel for helpful discussions. This paper makes use of restricted access data from the National Institutes of Health, protected by the Privacy Act of 1974 as amended (5 U.S.C. 552a). De-identified data necessary to reproduce all plots and statistical analyses will be made freely available. Those wishing to access the raw data may apply for access following the procedures outlined in the NIH Data Access Policy document available at 
{http://report.nih.gov/pdf/DataAccessPolicy.pdf}. The VentureXpert database is available via Thomson Reuters. The Global Terrorism Database is publicly available at 
{https://www.start.umd.edu/gtd/}. This work is supported by the Air Force Office of Scientific Research under award number FA9550-15-1-0162 and FA9550-17-1-0089, National Science Foundation grant SBE 1829344, and Northwestern University Data Science Initiative. This work does not reflect the position of NIH.
 \item[Competing Interests] The authors declare that they have no
competing financial interests.
 \item[Correspondence] Correspondence and requests for materials
should be addressed to D.W.\\(email: dashun.wang@northwestern.edu).
\end{addendum}

\newpage 

\section*{Figure captions}

\begin{figure}
    \centering
   \includegraphics[width=0.5\columnwidth]{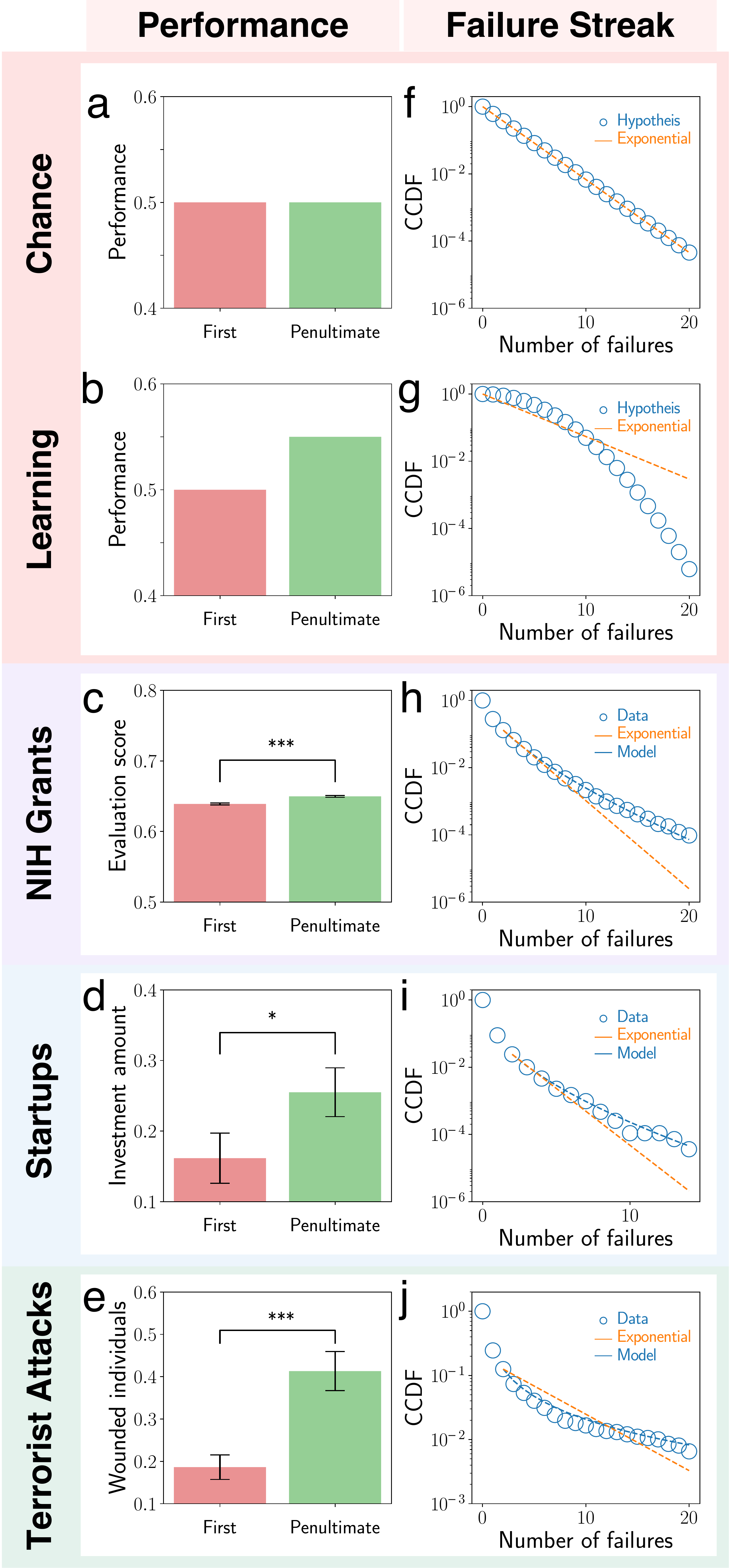}
  \caption{\textbf{The mechanisms of chance and learning.}}
 \label{hypothesis}
\end{figure}

\begin{figure}
    \centering
   \includegraphics[width=1\columnwidth]{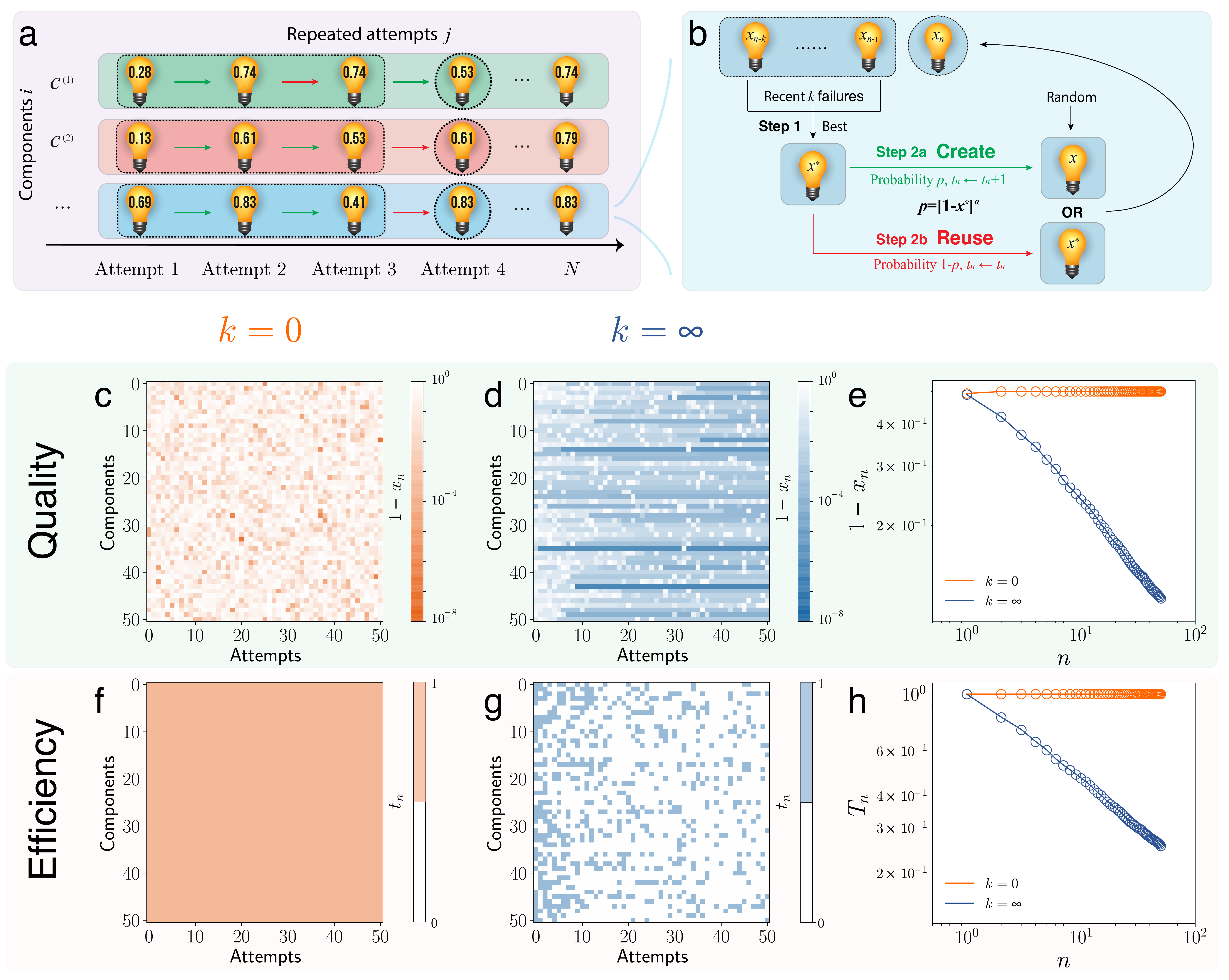}
  \caption{\textbf{The $k$ model.}}
 \label{scheme}
\end{figure}

\begin{figure}
    \centering
   \includegraphics[width=1\columnwidth]{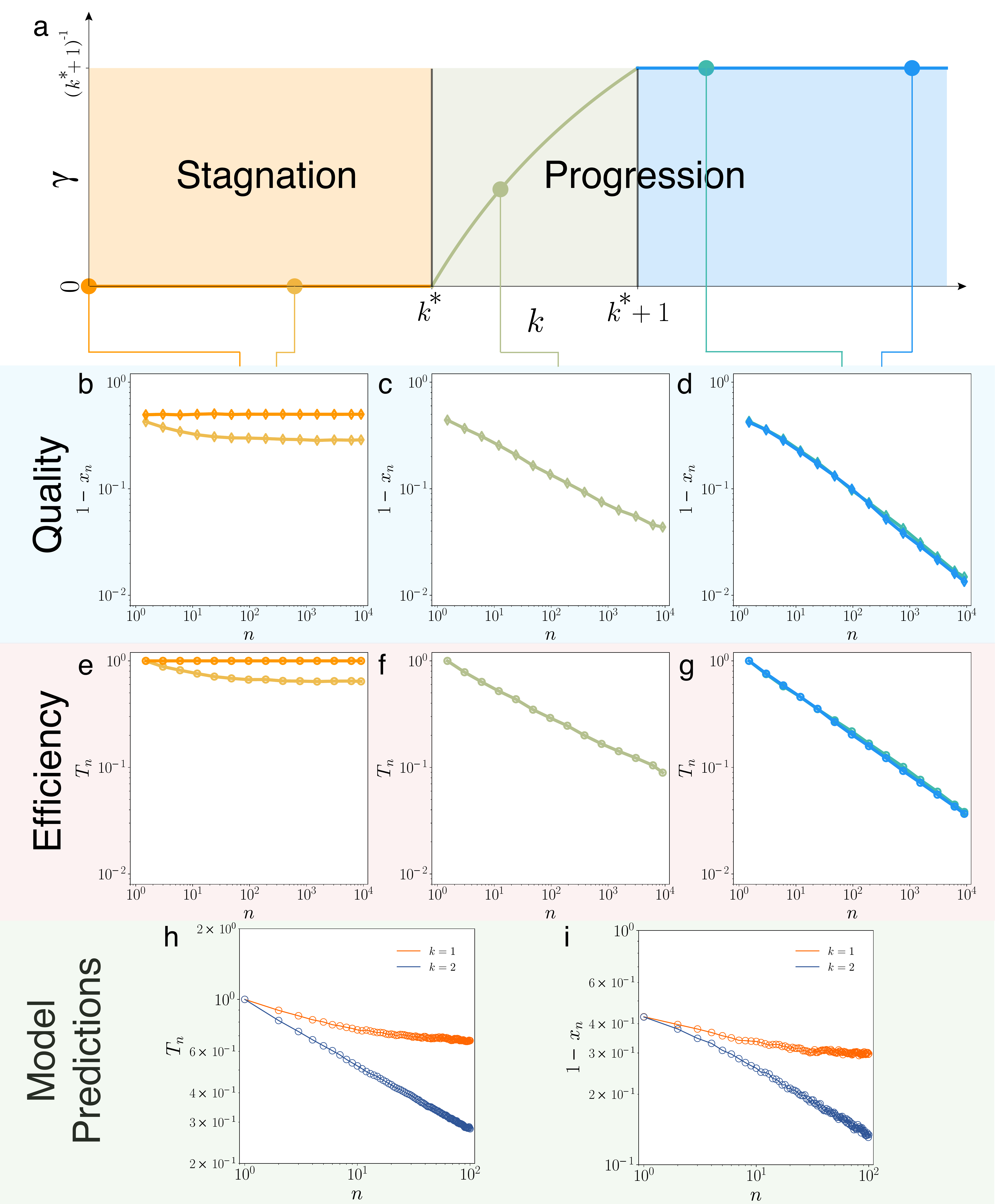}
  \caption{\textbf{Phase diagram of the model.}}
 \label{solution}
\end{figure}

\begin{figure}
    \centering
   \includegraphics[width=0.9\columnwidth]{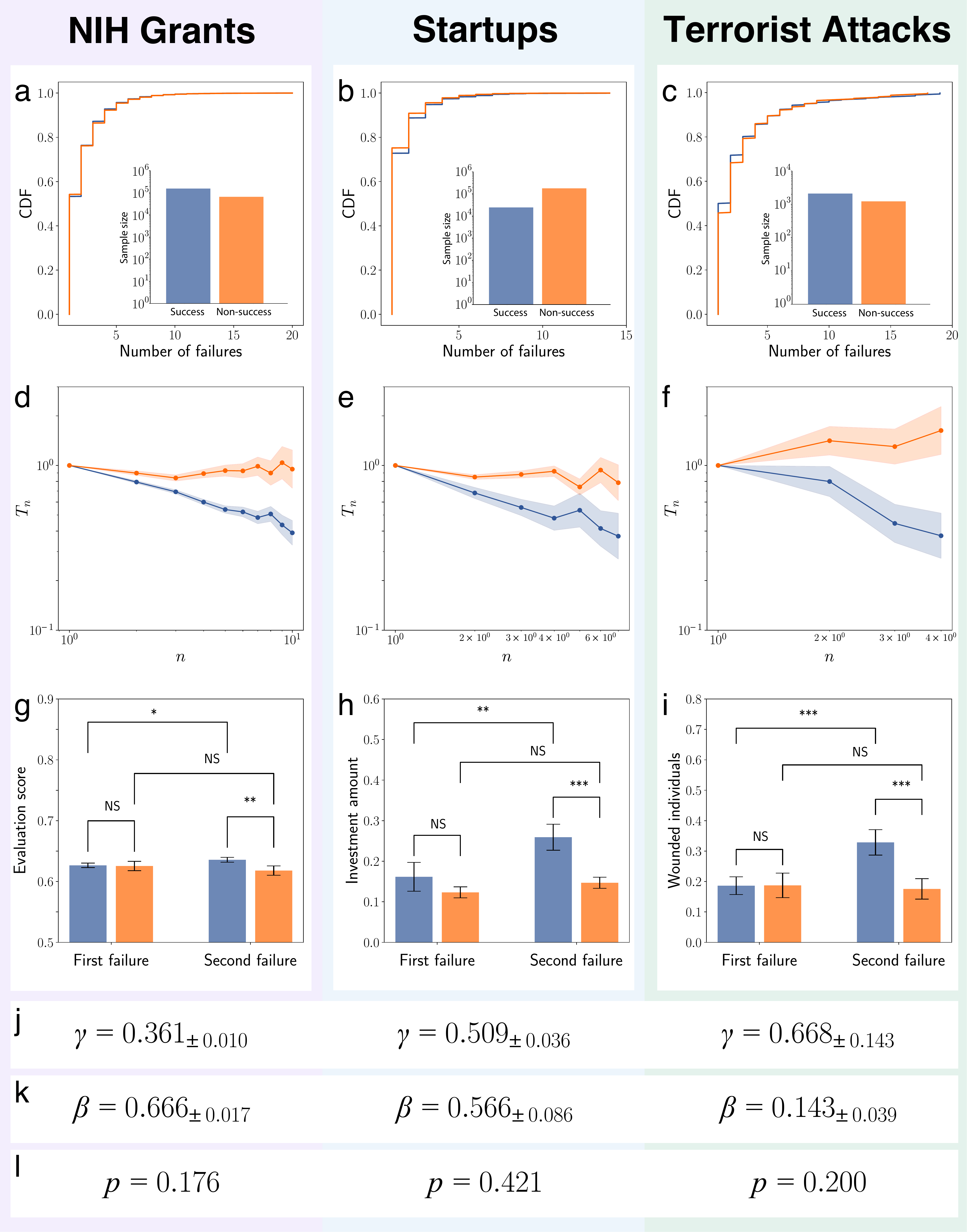}
  \caption{\textbf{Testing model predictions.}}
  \label{empirical}
\end{figure}

\textbf{Figure \ref{hypothesis}: The mechanisms of chance and learning.} We compare theoretical predictions and empirical measurements for performance changes ({\bf a-r}) as well as the length distribution of failure streaks ({\bf f-j}). The chance model predicts no performance change ({\bf a}), with failure streak length following an exponential distribution ({\bf f}). The learning hypothesis predicts improved performance ({\bf b}), with shorter failure streaks than expected by the chance model, corresponding to a faster-than-exponential distribution ({\bf g}). Both hypotheses are contested by empirical patterns observed across all three datasets. We measured the performance of an attempt based on NIH percentile scores ($D_1$), investment sizes ($D_2$), and number of wounded individuals ($D_3$). To ensure that performance metrics are comparable across data and models, we standardized performance measures according to their underlying distribution (SI S4.1). We find that failures in real data are characterized improved performance between the first and penultimate attempt ({\bf c-e}). Yet at the same time, failure streaks are characterized by a fat-tailed length distribution, indicating that failure streaks in real data are longer than expected by chance ({\bf h-j}).
\\
\textbf{Figure \ref{scheme}: The $k$ model.} ({\bf a}) 
Here we treat each attempt as a combination of independent components $(c^{(i)})$.
For an attempt $j$, each component $i$ is characterized by an evaluation score $x_j^{(i)}$, which falls between 0 and 1. The score for a new version is often unknown until attempted, hence a new version is assigned a score, drawn randomly from $[0,1]$, which approximates the percentile of any score distribution that real systems follow.
({\bf b}) To formulate a new attempt, one can either create a new version (with probability $p$, green arrow in {\bf a}), or reuse an existing version by choosing the best one among past versions $x^*$ (with probability $1-p$, red arrow in  {\bf a}). Reusing the existing best version allows the particular component to retain its score $x^*$ and avoids incurring additional time cost. Creating a new version costs one unit of time but generates a new score $x$.
Of the many factors that may influence $p$, one key factor is the quality of existing versions, suggesting that $p$ should be a function of $x^*$. Indeed, consider the two extreme cases. If $x^*\rightarrow 0$, existing versions of this component have among the worst scores and, hence, a high potential for improvement with a new version. Therefore the likelihood of creating a new version is high, i.e., $p\rightarrow 1$. On the other hand, $x^*\rightarrow 1$ corresponds to a near-perfect version, yielding a decreased incentive to create a new version ($p\rightarrow 0$). Therefore, $P(x\geq x^*)=1-x^*$ captures the potential to improve on prior versions, prompting us to assume $p=(1-x^*)^{\alpha}$, where $\alpha>0$ characterizes an agent's propensity to create new versions given the quality of existing ones. 
({\bf c-h}) Simulation results from the model ($\alpha=0.6$) for the cases of $k=0$ (c,f) and $k\rightarrow\infty$ (d,g) in terms of the average quality (c-e) and efficiency (f-h) of each attempt. $k=0$ recovers the chance model, predicting a constant quality (c) and efficiency (f). $k\rightarrow\infty$ predicts a temporal scaling characterizing the dynamics of failure (g) with an improved quality (d), recovering the predictions from learning curves and Wright's Law. 
\\
\textbf{Figure \ref{solution}: Phase diagram of the model.} {\bf (a)} Analytical solution of the model reveals that the system is separated into three regimes by two critical points $k^*$ and $k^*+1$. The solid line shows an extended solution space of our analytical results. {\bf (b-g)} Simulations results of the model ($\alpha=0.6$) for quality (b-d) and efficiency (e-g) trajectories for different $k$ parameters, showing distinctive dynamical behavior in different regions separated by the two critical points. All results are based on simulations over $10^4$ times. {\bf (h,i)} Phase transition around $k^*$ predicts the coexistence of stagnation ($k=1$, orange) and progression ($k=2$, blue) groups.
\\
\textbf{Figure \ref{empirical}: Testing model predictions.} 
\textbf{(a-c)} Complementary cumulative distribution (CCDF) of the number of consecutive failures prior to the last attempt for the success (blue) and non-success groups (orange). In each of our three datasets, two distributions are statistically indistinguishable (Kolmogorov-Smirnov test for samples with at least one failures). (Inset) The sample size of success and non-success group, showing their size is of a similar order of magnitude.
\textbf{(d-f)} Early temporal signals separate success and non-success groups. For each group we measure the average inter-event time between two failures $T_n\equiv t_n/t_1$ as a function of the number of attempts. Dots and shaded areas show the mean and standard errors of the mean measured from data (SI S4.3). All success groups manifest power law scaling $T_n\sim n^{-\gamma}$, with $\gamma$ reported in Table 1. This temporal scaling is absent for non-success groups.
\textbf{(g-i)} Performance at first attempt is indistinguishable between the success and non-success groups, but becomes distinguishable from the second attempt. Whereas performance improves for the success group, this improvement is absent for the non-success group.
\textbf{(j-l)} Parameter estimates (mean$_{\pm\text{standard error}}$). $\gamma$ corresponds to the temporal scaling exponent uncovered in (2) (j) and $\beta$ is the shape parameter of the Weibull distribution, characterizing the length distribution of failure streaks (k). Statistical tests indicate that none of the three datasets can reject the validity of the scaling identity $\beta+\gamma=1$ (l).

\clearpage
\includepdf[pages=1-89]{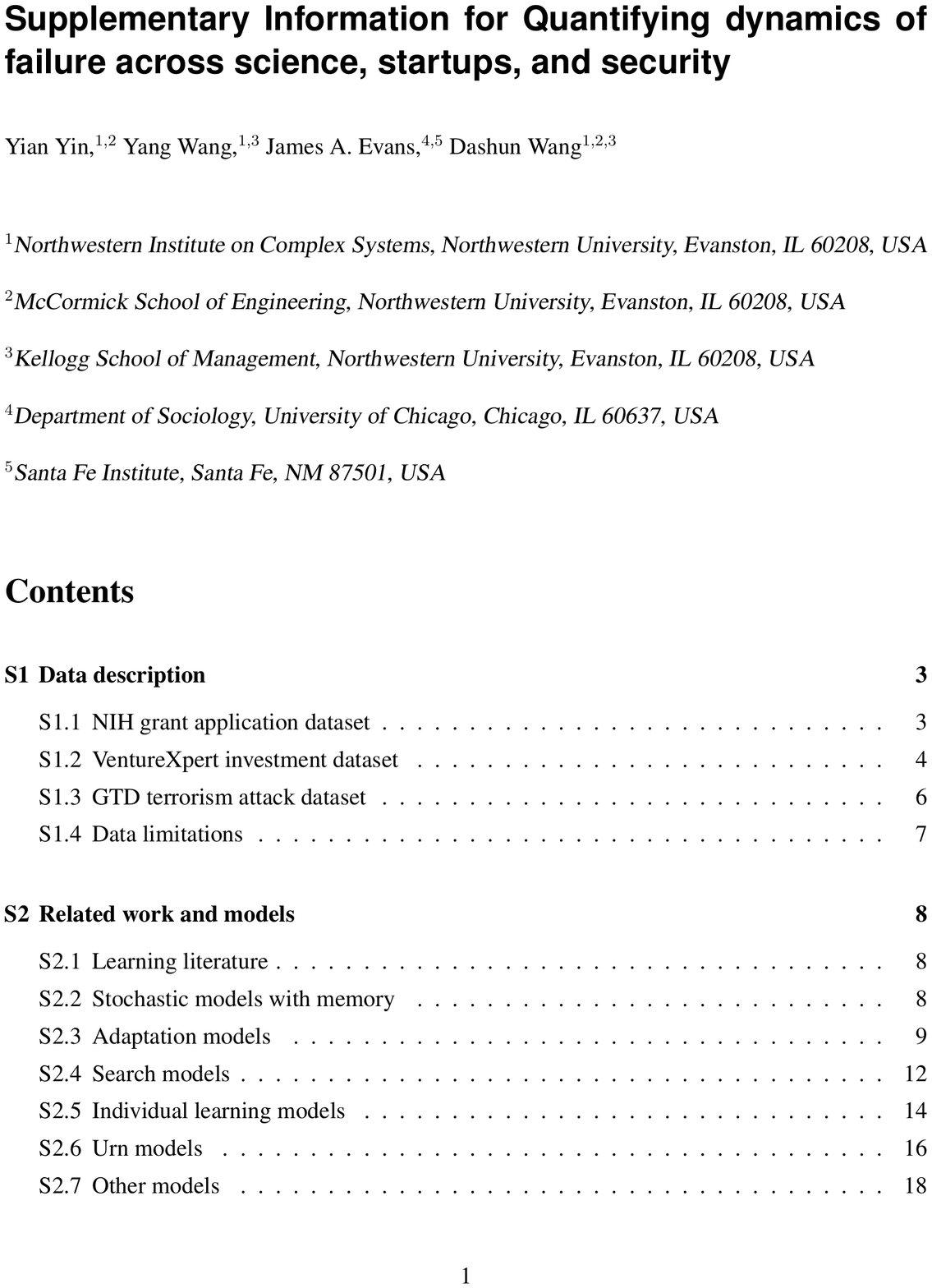}

\end{document}